\documentclass[aps,prb,twocolumn,showpacs]{revtex4-1}
\usepackage{mathptmx}

\usepackage[T1]{fontenc}
\usepackage[latin9]{inputenc}
\usepackage{float}
\usepackage{amsmath}
\usepackage{epsfig}
\usepackage{epstopdf}
\usepackage{amssymb}
\usepackage{graphicx}
\usepackage{esint}
\usepackage[normalem]{ulem}
\makeatletter
\@ifundefined{textcolor}{}
{%
\definecolor{BLACK}{gray}{0}
\definecolor{WHITE}{gray}{1}
\definecolor{RED}{rgb}{1,0,0}
\definecolor{GREEN}{rgb}{0,1,0}
\definecolor{BLUE}{rgb}{0,0,1}
\definecolor{CYAN}{cmyk}{1,0,0,0}
\definecolor{MAGENTA}{cmyk}{0,1,0,0}
\definecolor{YELLOW}{cmyk}{0,0,1,0}
}
\pdfoutput=1 
\usepackage{multirow}\usepackage{dcolumn}\usepackage{bm}\usepackage{ulem}\usepackage{amsfonts}\usepackage{color}\usepackage{colordvi}\usepackage{dcolumn}\usepackage{subfigure}
\usepackage[colorlinks,bookmarks=false,citecolor=blue,linkcolor=blue,urlcolor=blue,hyperfootnotes=true]{hyperref}
\makeatother
\begin{document}
\title{\textcolor{black}{$\eta$ collective mode as A$_{1g}$ Raman resonance in cuprate superconductors}}
\date{\today}
\author{X. Montiel} 
\email{xavier.montiel@cea.fr}
\affiliation{Institut de Physique Th\'{e}orique, CEA-Saclay, 91191 Gif-sur-Yvette, France}
\author{T. Kloss}
\email{thomas.kloss@cea.fr}
\affiliation{Institut de Physique Th\'{e}orique, CEA-Saclay, 91191 Gif-sur-Yvette, France}
\author{C. P\'{e}pin}
\email{catherine.pepin@cea.fr}
\affiliation{Institut de Physique Th\'{e}orique, CEA-Saclay, 91191 Gif-sur-Yvette, France}
\author{S. Benhabib}
\affiliation{Laboratoire Mat\'{e}riaux et Ph\'{e}nom\`{e}nes Quantiques (UMR 7162 CNRS), Universit\'{e} Paris Diderot-Paris 7, Bat. Condorcet, 75205 Paris Cedex 13, France}
\author{Y. Gallais}
\affiliation{Laboratoire Mat\'{e}riaux et Ph\'{e}nom\`{e}nes Quantiques (UMR 7162 CNRS), Universit\'{e} Paris Diderot-Paris 7, Bat. Condorcet, 75205 Paris Cedex 13, France}
\author{A. Sacuto}
\affiliation{Laboratoire Mat\'{e}riaux et Ph\'{e}nom\`{e}nes Quantiques (UMR 7162 CNRS), Universit\'{e} Paris Diderot-Paris 7, Bat. Condorcet, 75205 Paris Cedex 13, France}
\begin{abstract}
We discuss the possible existence a spin singlet excitation with charge $\pm2$ ($\eta$-mode) originating the $A_{1g}$ Raman resonance in cuprate superconductors. This $\eta$-mode relates the $d$-wave superconducting singlet pairing channel to a $d$-wave charge channel. We show that the $\eta$ boson forms a particle-particle bound state below the $2\Delta$ threshold of the particle-hole continuum where $\Delta$ is the maximum $d$-wave gap. Within a generalized random phase approximation and Bethe-Salpether approximation study, we find that this mode has energies similar to the resonance observed by Inelastic Neutron Scattering (INS) below the superconducting (SC) coherent peak at $2\Delta$ in various SC cuprates compounds. We show that it is a very good candidate for the resonance observed in Raman scattering below the $2\Delta$ peak in the $A_{1g}$ symmetry. Since the $\eta$-mode sits in the $S=0$ channel, it may be observable via Raman, X -ray or Electron Energy Loss Spectroscopy probes.
\end{abstract}
\maketitle
\begin{section}{Introduction}

\textcolor{black}{The study of collective excitations of a condensed matter system can provide informations about the symmetries of an order parameter or the nature of the interactions between particles. The analysis of collective modes can also help  us to understand the origin of high critical temperature superconductivity (SC) in cuprate compounds (see schematic cuprate compounds  phase diagram on Fig. \ref{phasediagram}).}

\textcolor{black}{The existence of a collective spin excitation in the SC state of cuprates has been highlighted by Inelastic Neutron scattering (INS) experiments with the observation of a resonance at $41\,$meV in YBCO compounds and at similar energies in other compounds \cite{Rossat,Bourges,Hinkov04} around the antiferromagnetic (AF) ordering vector $\bf{Q}$$=(\pi,\pi)$. This resonance, which stands below the $2\Delta_{SC}^{0}$ threshold of the particle-hole continuum ($\Delta_{SC}^{0}$ is the maximum of the $d$-wave SC gap), seems to scale with the superconducting gap energy above optimal doping (at least until $p=0.19$ hole doping in Bi2212) \cite{Sidis2001}.}

\textcolor{black}{This INS resonance has initially been explained in the framework of the SO(5) emergent symmetry model for cuprates as a $\pi$-mode \cite{Demler95,Demler04}. However this $\pi$-mode lies at higher energies than experimentally observed, and the neutron resonance is now explained as a spin triplet exciton or resonance which emerges in the SC state because of a residual spin-spin interaction in the system \cite{NormanChub01} (see also Ref. \cite{Pepin94}).}

\textcolor{black}{A resonance very similar to the Neutron resonance has also been observed in Raman scattering experiments in the $A_{1g}$ symmetry channel \cite{Cooper88,Staufer92,Sacuto97,Gallais04,LeTacon05b}. In YBCO the A$_{1g}$ Raman resonance is located at $41$ meV at optimal doping, and follows the neutron resonance energy with nickel and zinc substitutions \cite{SacutoSidis02,LeTacon06b}. 
This resonance is not seen in the $B_{1g}$ symmetry, which is scanning the anti-nodal region of the Fermi surface, nor in the $B_{2g}$ channel, which is scanning the nodal region. In the $B_{1g}$ channel, a SC coherence peak is observed at higher energy than the $A_{1g}$ resonance. Its energy matches well twice the maximum of the $d$-wave SC gap $2\Delta_{SC}^{0}$ observed in other spectroscopies.}

\textcolor{black}{The $A_{1g}$ Raman resonance has initially been attributed to the $2\Delta_{SC}^{0}$ SC pairing gap. However, later considerations
showed that long range Coulomb screening washes out the single particle pair-breaking contribution to the Raman response in the $A_{1g}$ channel leaving the position and the intensity of the $A_{1g}$ Raman resonance essentially unexplained \cite{Devereaux95, cardona97, Venturini2000, Devereaux-RMP}. Note that the two-magnons process proposed in Ref. \cite{Venturini2000} produces a resonance at twice the energy of the neutron resonance and thus cannot explain the near perfect energy matching between the $A_{1g}$ Raman resonance and the Neutron resonance around the optimal doping.}

\textcolor{black}{From a theoretical point of view there are several open questions regarding the Raman resonance in the $A_{1g}$ channel :
Since the $A_{1g}$ symmetry scans the whole Brillouin Zone (BZ), how can we explain the absence of a superconducting coherent peak in the $A_{1g}$ channel whereas it is observed in the $B_{1g}$ channel? How can we explain that the $A_{1g}$ resonance emerges at lower energy than the SC coherent peak one? How can we explain the perfect match between the energy of the Raman resonance in the $A_{1g}$ channel and the INS frequency resonance at optimal doping? What are the informations that this $A_{1g}$ Raman resonance could give us about the physics of Cuprates ?}

\textcolor{black}{In this paper, we present a coherent scenario that provides an answer to the latter questions. Our scenario builds on recent experimental developments \cite{Hoffman02,Hanaguri04,DoironLeyraud07,Sebastian12,Doiron-Leyraud13,Wise08,Fujita12,He14,Tabis14,Blanco-Canosa14,Vishik12} that pinpointed the existence of a Charge Density Wave (CDW) state, with ordering vector $\bf{q_{c}}$$=(q_{x},0)$ and $\bf{q_{c}}$$=(0,q_{y})$ with $q_{x}=q_{y}\approx0.3\pi$ which exists between hole doping $p\approx0.1$ until $p\approx0.19$ \cite{Fujita14}  (see Fig. \ref{phasediagram}, note that the endpoint of the CDW order is still under debate). 
We explain the Raman resonance in the $A_{1g}$ symmetry as a collective mode that allows excitations between the $d$-wave SC pairing sector and the observed $d$-wave CDW charge sector. The proposed scenario is supported by the fact that the $A_{1g}$ Raman resonance exists between $p=0.12$ and $p=0.22$ \cite{Sacuto2015TS} which overlaps with the hole doping range where a coexistence between a $d$-wave SC and a $d$-wave charge density wave (CDW) ordered states have been observed \cite{Fujita14} (see figure \ref{phasediagram}). This collective mode is a spin singlet $(S=0)$ excitation with a charge $\pm2$ and we call it a $\eta$-mode of the system.}

\begin{figure}
\includegraphics{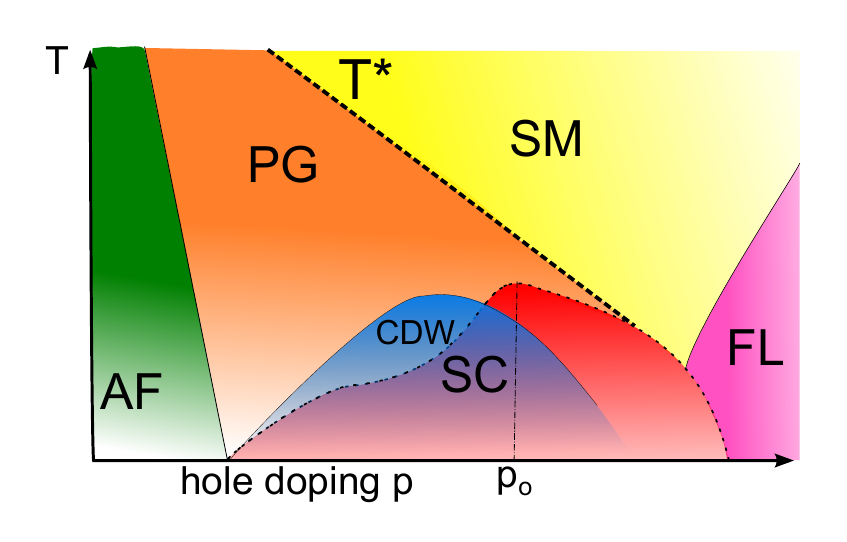}
\caption{\label{phasediagram}(Color online) Schematic Temperature-hole doping (T,p) phase diagram of superconducting cuprates. Close to zero hole doping, the system exhibits a Mott insulator antiferromagnetic phase (AF). At higher hole doping, the Pseudo-Gap (PG) phase appears below the critical temperature $T^{\star}$ (dashed line). A dome of a $d$-wave superconducting (SC) phase appears at average doping below the critical temperature $T_{c}$ (dotted line). The maximum of critical temperature $T_{c}$ is obtained at the optimal doping $p_{o}$. Above $T^{\star}$ at average doping, the cuprate compounds exhibits a metallic phase with a linear temperature dependence of the resistivity called Strange metal (SM). The conventional Fermi liquid (FL) state appears at high doping. A charge density wave (CDW) order appears below $T^{\star}$ and above $T_{c}$ (blue area). The hole doping where the CDW state disappear is very discussed then we do not finish the transition line. Note that $T^{\star}$ meet $T_{c}$ around $p=0.22$ in Bi2212 compound only. In other compounds, the behaviour of the $T^{\star}$ line can be different.}
\end{figure}

In order to explain the $A_{1g}$ Raman resonance, we assume the coexistence between a $d$-wave SC phase and a $d$-wave CDW ordered phase. The CDW phase has the same ordering vector as the observed one $\bf{q_{c}}$$=(q_{x},0)$ and $\bf{q_{c}}$$=(0,q_{y})$ with $q_{x}=q_{y}\approx0.3\pi$. This coexisting phase has been observed by STM in Bi2212 compound \cite{Fujita14}.

We note that in the references \cite{Zeyher99,Zeyher02}, the authors develops an effective $t-J$ model with coexistent SC and CDW order that reproduce qualitatively the dependence of the frequency resonance in the underdoped phase. However, they considered a charge ordering wave-vector $\bf{Q}=(\pi,\pi)$, which is not observed experimentally and did not reproduce the peak in the $B_{1g}$ symmetry.

Here, we argue that the $A_{1g}$ Raman \textcolor{black}{resonance is the collective mode describing the proximity of the charge and pairing channels. Fluctuations between those two sectors are typically described by non-linear $\sigma$- models as in Refs \cite{Hayward14,Efetov13}.}
Within a model of itinerant electrons interacting through an effective AF spin-spin coupling, we find that the $\eta$-resonance forms a particle-particle bound state situated below the $2\Delta_0$-threshold, in a very similar way to the triplet spin exciton revealed by neutron scattering.
 
The paper is organized as follows: in the section \ref{theor}, we present the theoretical model we use to reproduce the experimental Raman data. We explain how we describe the different phases of the Cuprates (section \ref{description}). We present the different susceptibilities we consider (section \ref{Susceptibilities}) and give details about the evaluation of the different Feynman diagrams (section \ref{Eval}). We also explain the calculations of the full Raman response (section \ref{Raman}). In the section \ref{results}, we present the results of our theory and we compare them to the experimental data before to conclude (section \ref{conclusion}). Explanations and information about the spin fermion model and the experimental setup are given in details in the appendix \ref{Frequency} and \ref{RamanExperiment} respectively.

\end{section}
\begin{section}{Theoretical model}
\label{theor}
\textcolor{black}{A global explanation of the phase diagram of cuprates compounds is still lacking. For example, we still do not know whether the cuprates are fundamentally doped Mott insulators admitting a Coulomb energy repulsion of 1eV which is crucial to explain the emergence of superconductivity (see e.g\ \cite{Lee06,Gull:2013hh,Sorella02}), or whether an itinerant electron picture with strong Anti-Ferromagnetic (AF) fluctuations is a good approximation to explain the main features of the phase diagram (e.g.\ \cite{abanov03,Norman03,sfbook}). A consensus exists in the recognition that three main players are present in the phase diagram: $d$-wave superconductivity, AF order and fluctuations, and Mott insulating phase. Note that the charge order could be a fourth key player of the cuprate compounds physics.} 

\textcolor{black}{Below $T^{*}$, the PG phase critical temperature \cite{Alloul89,Warren89,Tallon200153} (see Fig. \ref{phasediagram}), the influence of charge orderings seems to be stronger than previously thought. The proximity between pairing and charge channel is well described by spin fermion approach. Such approach have been used  through simplified Eight Hot Spots (EHS) models in order to explain the PG phase \cite{Metlitski10b,Efetov13}. In Ref.\cite{Efetov13}, a coexisting $d$-wave bond order and $d$-wave  SC state are related by SU(2) symmetry. Note that the SU(2) symmetry relating the charge and spin channels is only realized in a strict sense in the eight hot spots model where the Fermi dispersion has been linearized around each hot spot.}

\textcolor{black}{ However, the CDW and PG phases are different phases as the CDW phase appears on the tips of the Fermi arcs after the antinodal zone has been gaped out (see Fig. \ref{phasediagram}). The study of SC and CDW competing orders related by angular fluctuations was performed in a model based on Landau theory\cite{Hayward14}.
In this model, the SU(2) symmetry can be recovered by ignoring the quartic and anisotropic couplings.  Note that in Ref. \cite{Wang14}, a spin fermion approach was developped to explain the stability of CDW phase related to modulated SC order called pair density wave (PDW) which has not been observed experimentally. In the following, we leave aside the complexity of the spin fermion approach and we describe the SC/CDW coexisting phases with another model.}

\begin{subsection}{Description of the system}
\label{description}
\begin{subsubsection}{The two body Hamiltonian}
We consider a system of itinerant fermions interacting through an effective AF spin-spin coupling close to vector $\mathbf{Q}$ derived from spin-fermion approach \cite{Efetov13} (detailed in Appendix \ref{Frequency}). \textcolor{black}{We then describe a system of fermions with a kinetic energy and nearest-neighbor (n.n.) interaction Hamiltonian involving both spin and charge in the weak coupling regime.} The Hamiltonian writes $H= \sum_{i,j,\sigma}t_{ij}c_{i\sigma}^{\dagger}c_{j\sigma}+\sum_{\left\langle i,j\right\rangle }\left(J_{ij}\mathbf{S}_{i}\cdot\mathbf{S}_{j}+V_{ij} n_{i}n_{j}\right)$ where $n_{i}=\sum_{\sigma}c_{i\sigma}^{\dagger}c_{i\sigma}$ and $\mathbf{S}_{i}=\sum_{\alpha \beta}c_{i\alpha}^{\dagger}\mathbf{\sigma_{\alpha,\beta}}c_{i\beta}$ are the density and spin operators respectively with $\sigma_{\alpha\beta}$ the Pauli matrix vector. $\langle i j \rangle$ denotes summation over nearest neighbors and $t_{ij}$ is the hopping parameter. $J_{ij}$ is the n. n. super exchange coupling and $V_{ij}$ the n.n. Coulomb term. Long-range Coulomb effects will be considered later while discussing the Raman response. We will neglect them for the study of the collective mode. 
Involving the Fourier transform $c_{i,\sigma}=\frac{1}{\sqrt{N}}\sum_{\mathbf{k}}e^{i \mathbf{k} \cdot \mathbf{r}_{i}}c_{\mathbf{k},\sigma}$, where $N$ is the total number of lattice sites, the Hamiltonian writes:
\begin{align}
H= & \sum_{\mathbf{k},\sigma}\xi_{\mathbf{k}}c_{\mathbf{k}\sigma}^{\dagger}c_{\mathbf{k}\sigma}\nonumber\\
&+\sum_{\mathbf{k,k',q}}\left(J_{\mathbf{q}}c^{\dagger}_{\mathbf{k},\alpha}\sigma^{T}_{\alpha \beta}c_{\mathbf{k+q},\beta}c^{\dagger}_{\mathbf{k'+q},\gamma}\sigma_{\gamma \delta}c_{\mathbf{k'},\delta}\right) \nonumber\\
&+\sum_{\mathbf{k},\mathbf{k'},\sigma,\sigma'}V_{\mathbf{q}}c^{\dagger}_{\mathbf{k},\sigma}c_{\mathbf{k+q},\sigma}c^{\dagger}_{\mathbf{k'},\sigma'}c_{\mathbf{k'-q},\sigma'}
,\label{eq:1}
\end{align}
where $c^{(\dagger)}_{\mathbf{p},\alpha}$ is the annihilation (creation) operator of an electron with spin $\alpha$ and impulsion $\mathbf{p}$. $\xi_{\mathbf{k}}$  is the one particle energy with a tight-binding form obtained to fit ARPES data \cite{Kordyuk2003} : $\xi_{\mathbf{k}}=-2t(\cos(k_{x}a)+\cos(k_{y}a))+4t'\cos(k_{x}a)\cos(k_{y}a)+t_{0}(\cos(k_{x}a)-\cos(k_{y}a))^{2}-\mu$ where $t,t'$ are respectively the first and second neighbor hopping terms with $t'=-0.3t$ and $t_{0}=0.084t$  which is third and fourth neighbor hopping term. $a$  is the cell parameter set to unity and $\mu$ is the chemical potential determined to adjust the hole doping.
$J_{\mathbf{q}}=-\frac{J}{2}\left(\cos(q_{x}a)+\cos(q_{y}a)\right)$ is the Fourier transform of $J_{ij}$ developed around $\mathbf{Q}$, while $V_{\mathbf{q}}=\frac{V}{2}\left(\cos(q_{x}a)+\cos(q_{y}a)\right)$ denotes the amplitude of the nearest-neighbour Coulomb interaction for small values of $\mathbf{q}$. \textcolor{black}{$V_{\mathbf{q}}$ accounts for repulsion between charge and vanishes in conventional $t-J$ model. In the following, we treat $V_{\mathbf{q}}$ and $J_{\mathbf{q}}$ as independent parameters \cite{Demler95,Schrieffer64}.}
\end{subsubsection}
\begin{subsubsection}{Order parameters and effective Hamiltonian}
In the following, we focus on the superconducting phase where the Raman $A_{1g}$ resonance peak has been observed. We consider that the superconducting phase around optimal doping is a coexistence between a $d$-wave SC order and the $d$-wave CDW order. Thus, we decouple the Hamiltonian (\ref{eq:1}) in the SC pairing channel and in the charge channel by introducing the two order parameters $\Delta_{SC,\bf{k}}$ and $\Delta_{CDW,\bf{k}}$.
$\Delta_{SC,\bf{k}}$ is the superconducting order parameter which describes the $d$-wave SC order. $\Delta_{CDW,\bf{k}}$ is the CDW order parameter which describes the $d$-wave charge order with ordering vector $\bf{q_{c}}$.
Applying the Hubbard-Stratanovitch transformation to the Hamiltonian (\ref{eq:1}), the effective Hamiltonian of the system can be written in the basis $\Psi^{\dagger}_{\bf{k}}=\left( c_{\bf{k},\sigma}^{\dagger},c_{\bf{-k-q_{c}},\overline{\sigma}},c_{\bf{k+q_{c}},\sigma}^{\dagger},c_{-\bf{k},\overline{\sigma}}\right) $
as 
\begin{align}
&\hat{G}^{-1}\left(\bf{k},\omega\right)=\left(\begin{array}{cccc}
i\omega-\xi_{\bf{k}} & 0 & \Delta_{CDW,\bf{k}} & \Delta_{SC,\bf{k}}\\
0 & i\omega+\xi_{\bf{-k-q_{c}}} & \Delta_{SC,\bf{k+q_{c}}}^{\dagger} & -\Delta_{CDW,\bf{k}}\\
\Delta_{CDW,\bf{k}}^{\dagger} & \Delta_{SC,\bf{k+q_{c}}} & i\omega-\xi_{\bf{k+q_{c}}} & 0\\
\Delta_{SC,\bf{k}}^{\dagger} & -\Delta_{CDW,\bf{k}}^{\dagger} & 0 & i\omega+\xi_{-\bf{k}}
\end{array}\right)\label{eq:green-1}
\end{align}
with $\xi_{-\bf{k}}=\xi_{\bf{k}}$. Note that the Green function inverse matrix is very similar to the one calculated in \cite{Zeyher99,Zeyher02}.
From inverting the matrix ($\ref{eq:green-1}$) , one can find the Green
functions of the system. Particularly, the CDW quasiparticle Green function $G^{\Delta_{CDW}}(\bf{k},\omega)$ and the anomal SC Green function $G^{\Delta_{SC}}(\bf{k},\omega)$ writes :
\begin{align}
&G^{\Delta_{CDW}}(\textbf{k},\omega)=\Delta_{CDW,\bf{k}} \times \nonumber\\
&\frac{\left(\Delta_{SC,\bf{k+q_{c}}}^{\dagger}\Delta_{SC,\bf{k}}+\left|\Delta_{CDW,\bf{k}}\right|^{2}-\xi_{\bf{k+q_{c}}}\xi_{k}+\omega^{2}-i\omega(\xi_{\bf{k+q_{c}}}+\xi_{\bf{k}})\right)}{\text{det}\left(\hat{G}\left(\bf{k}, \omega\right)\right)}\label{GCDW}\\
&G^{\Delta_{SC}}(\bf{k},\omega)\nonumber\\
&=\frac{\left(\omega^{2}+\xi_{\bf{k+q_{c}}}^{2}+\left|\Delta_{SC,\bf{k+q_{c}}}\right|^{2}\right)\Delta_{SC,\bf{k}}+\Delta_{SC,\bf{k+q_{c}}}\left|\Delta_{CDW,\bf{k}}\right|^{2}}{\text{det}\left(\hat{G}\left(\bf{k},\omega\right)\right)}\label{GSC}
\end{align}
with 
\begin{align}
&\text{det}\left(\hat{G}^{-1}\left(\bf{k},\omega\right)\right)=\left|\Delta_{CDW,\bf{k}}\right|^{4}\nonumber\\
&+2\left|\Delta_{CDW,\bf{k}}\right|^{2}\left(\Delta_{SC,\bf{k}}\Delta_{SC,\bf{k+q_{c}}}+\omega^{2}-\xi_{\bf{k+q_{c}}}\xi_{\bf{k}}\right)\nonumber\\
&+\left(\omega^{2}+\xi_{\bf{k}}^{2}+\left|\Delta_{SC,\bf{k}}\right|^{2}\right)\left(\omega^{2}+\xi_{\bf{k+q_{c}}}^{2}+\left|\Delta_{SC,\bf{k+q_{c}}}\right|^{2}\right)
\label{eq:det-1}.
\end{align}
The Green functions of the system will be crucial to estimate the different susceptibilities of the system (see section \ref{Susceptibilities}).
\end{subsubsection}
\begin{subsubsection}{\textcolor{black}{Phenomenology and symmetries of the SC and CDW orders}}
\label{sym}
In principle, we have to solve the self-consistent equations to determine $\Delta_{SC}(k,\omega)$ and $\Delta_{CDW}(k,\omega)$. However, the goal of this paper is not to explain the (T,p) phase diagram. Based on experimental evidence that the charge is modulated inside the superconducting phase of cuprates \cite{Fujita14},we assume that the SC phase is composed by a competition between a $d$-wave SC phase and a $d$-wave CDW order. The amplitude of the order parameters $\Delta_{SC,\bf{k}}$ and $\Delta_{CDW,\bf{k}}$ is expected to vary with the hole doping. The CDW order has the same ordering wave-vector $\bf{q_{c}}$ as the one observed experimentally. \textcolor{black}{In this paper, we do not address the problem of the PG state. We then neglect the PG energy scale and we assume that the PG state does not infer with the Raman response in the SC state. This assumption is expected to be valid around optimal doping and in the overdoped regime.}

In Bi2212, the SC order parameter decreases above optimal doping $p=0.16$ and vanishes around doping $p=0.25$ as mentioned in \cite{Fujita14}. The CDW order is maximal at doping $p=0.12$ and vanishes around $p=0.19$ at zero temperature \cite{Fujita14}, while the PG transition line meet the superconducting transition line at a slightly higher doping, around $p=0.22$ \cite{Vishik2012,sacuto15}.

We assume a simple momentum and frequency dependence of the SC pairing $\Delta_{SC} (k,\omega)= \frac{\Delta_{SC}^{0}}{2}\left(\cos(k_{x}a)-\cos(k_{y}a)\right)f(\omega,\Delta^{0}_{SC})$ and CDW order $\Delta_{CDW} (k,\omega)= \frac{\Delta^{0}_{CDW}}{2}\left(\cos(k_{x}a)-\cos(k_{y}a)\right)f(\omega,\Delta^{0}_{CDW})$ where $\Delta^{0}_{CDW} (\Delta_{SC}^{0})$ is the maximum of the $d$-wave CDW (SC) order and $f(\omega,X)=e^{-\frac{\omega^{2}}{2\sigma^{2}}}$ with a variance $\sigma=X/1.177$ ensuring a half width at half maximum equals to $X$. \textcolor{black}{Recent STM experiments demonstrate the $d$-wave character of CDW order \cite{Hamidian15}.}

\textcolor{black}{Above optimal doping, the phase diagram (fig \ref{phasediagram}) shows that the superconducting state disappears at an higher temperature than the CDW state and we can consider that the CDW gap amplitude is lower than the SC one :  $\Delta^{0}_{SC}>\Delta^{0}_{CDW}$. In the underdoped regime, the decreasing of $T_{c}$ which goes below the CDW critical temperature (see fig \ref{phasediagram}) can be described by the opposite regime $\Delta^{0}_{SC}\le\Delta^{0}_{CDW}$.}

\textcolor{black}{The frequency dependence of the order parameters is a consequence of the spin-fermion model as studied in ref.\cite{Efetov13,Kloss151} and in Appendix \ref{Frequency}. We choose a frequency dependence of the order parameters that is a good approximation of the one coming from the spin fermion model (see Appendix \ref{Frequency}).}

\end{subsubsection}
\begin{subsubsection}{\textcolor{black}{SU(2) symmetry between SC and CDW order}}
\textcolor{black}{The concept of emerging symmetries has first been introduced to explain PG phase of cuprates compounds. Particularly, in Ref. \cite{Yang89,Yang90}, a representation with a pseudo-spin operators was introduced which rotates the $d$-wave SC state onto a charge order with ordering vector $q_{c}=(\pi,\pi)$.  In systems with coexisting $d$-wave SC and CDW states, one can define a pseudo-spin respecting the SU(2) symmetry which rotates from SC state onto a CDW order \cite{KeeSU2}.
Indeed, in $d$-wave SC and CDW coexisting orders, we can define a pseudo-spin where each components is a SC or CDW operator with $\Delta_{0}=\Delta_{CDW}$, $\Delta_{1}=\Delta_{SC}^{\dagger}$and $\Delta_{-1}=-\Delta_{1}^{\dagger},$}
\begin{subequations}
\label{SU2state}
\begin{align}
\Delta_{1} & =-\frac{1}{\sqrt{2}}\sum_{\mathbf{k}}d_{\mathbf{k}}c_{\mathbf{k}\uparrow}^{\dagger}c_{-\mathbf{k}\downarrow}^{\dagger},\\
\Delta_{0} & =\frac{1}{2}\sum_{\mathbf{k},\sigma}d_{\mathbf{k}}c_{\mathbf{k}\sigma}^{\dagger}c_{\mathbf{k+q_{c}}\sigma},\\
\Delta_{-1} & =\frac{1}{\sqrt{2}}\sum_{\mathbf{k}}d_{\mathbf{k}}c_{\mathbf{k}\downarrow}c_{-\mathbf{k}\uparrow}.
\end{align}
\end{subequations}
\textcolor{black}{The lowering and raising pseudo-spin operators $\eta^{+},$ $\eta^{-}=\left(\eta^{+}\right)^{\dagger}$, and $\eta_{z}$,  follow the definition :} 
\begin{subequations}
\label{SU2ope}
\begin{align}
\eta^{+} & =\sum_{\mathbf{k}}c_{\mathbf{k}\uparrow}^{\dagger}c_{-\mathbf{\mathbf{k}+q_{c}}\downarrow}^{\dagger}\\
\eta_{z} & =\sum_{\mathbf{k}}\left(c_{\mathbf{k}\uparrow}^{\dagger}c_{\mathbf{k}\uparrow}+c_{\mathbf{k+q_{c}}\downarrow}^{\dagger}c_{\mathbf{k+q_{c}}\downarrow}-1\right).
\end{align}
\end{subequations}
\textcolor{black}{The operators (\ref{SU2ope} form an SU(2) Lie Algebra. The SU(2) spin operators (\ref{SU2ope}) rotate each component of the pseudo-spin (\ref{SU2state}) into another \cite{KeeSU2}.}

\textcolor{black}{Typical approach to study this underlying SU(2) symmetry are non-linear $\sigma$ models \cite{Efetov13,Hayward14}. Such models have been proposed to explain the physics of cuprate superconductors \cite{Metlitski10b,Efetov13,Wang14,Hayward14}. In the following, we calculate the response of the operators (\ref{SU2ope}a) that connect SC and CDW orders.}
\end{subsubsection}
\end{subsection}
\begin{subsection}{Susceptibilities of the system}
\label{Susceptibilities}
\begin{subsubsection}{The collective modes}
In order to study the collective mode physics, we analyze the linear response for the two following operators. First, the INS resonance originates the excitation from the SC singlet state to the spin triplet at wave vector $\textbf{Q}$ associated with the AF order \cite{NormanChub01}. It is described by the operator
\begin{equation}
S_{+}=N^{-1/2}\sum_{\mathbf{k}}c_{\mathbf{k}\uparrow}^{\dagger}c_{\mathbf{k}+\mathbf{Q}\downarrow}.\label{eq:2bis}
\end{equation}
which destroy a bosonic excitation at momentum $\textbf{Q}$ with a $0$ charge and spin 1 as presented in Fig. \ref{vertexbis} d). 
The $\eta$ mode excites from the SC order parameter $\langle c_{\mathbf{k}\uparrow}^{\dagger}c_{\mathbf{-k}\downarrow}^{\dagger}\rangle$ to the CDW order $\langle c_{\mathbf{k}\uparrow}^{\dagger}c_{\mathbf{k+q_{c}}\uparrow}\rangle$. 
 It is described by the operator:
\begin{align}
\eta & =N^{-1/2}\sum_{\mathbf{k}}c_{\mathbf{k+q_{c}}\uparrow}c_{\mathbf{-k}\downarrow},\label{eq:2}
\end{align}
where $\mathbf{q_{c}}$ is the ordering vector of the charge density wave (CDW) order. The $\eta$ operator destroy a bosonic excitation at momentum $\mathbf{q_{c}}$ with a 0 spin, a $-2$ charge and is fully symmetric  as presented in Fig. \ref{vertexbis} e). Upon action of it, the SC state transforms into the CDW state in the same way than the operators (\ref{SU2ope}a).  The $\eta$-mode is obtained at wave vector $\mathbf{q_{c}}$ with $\mathbf{q_{c}}=(0,q_{y})=(q_{x},0)$ and $q_{x}=q_{y}=0.3\pi$, contrasting with the typical $\mathbf{Q}=\left(\pi,\pi\right)$ location of the spin triplet exciton. \textcolor{black}{Within the framework of the eight hot spots spin fermion model, where the electronic dispersion has been linearized around the Fermi points, the operator described in Eqn.(\ref{eq:2}) represents an SU(2) rotation between the CDW and SC states.}
\end{subsubsection}
\begin{subsubsection}{Susceptibilities in the Random Phase Approximation (RPA) and Bethe-Salpether Approximation}
\begin{figure}
\includegraphics{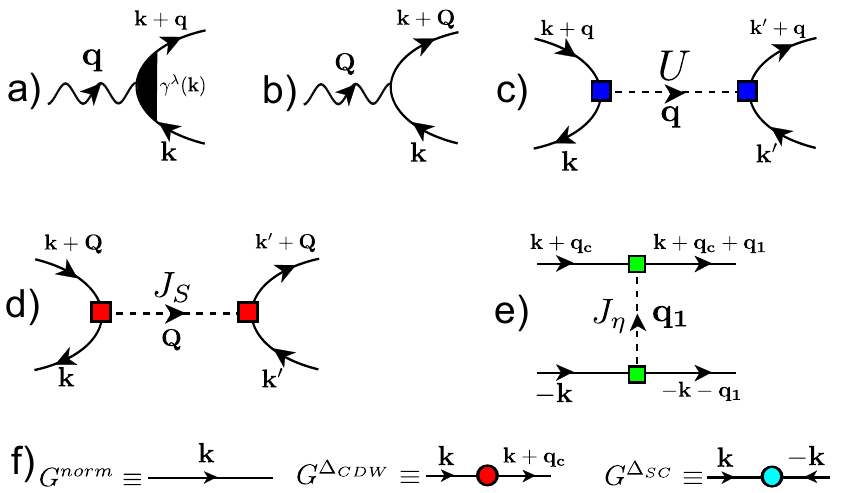}\vspace{-2.5ex}
\caption{\label{vertexbis}(Color online) Graphical representation for a) the photon-electron vertex at $\bf{q}$ and for the symmetry $\lambda$ and b) the neutron-spin vertex at $\bf{Q}=(\pi,\pi)$.  Graphical representation for a) the Coulombian interaction vertex (see ref. \cite{ cardona97} for details), b) the spin triplet exciton vertex and c) the $\eta$-mode vertex. In f) are presented the diagrammatic representation of the normal Green functions, the CDW Green function and the SC anomal Green function.}
\end{figure}
\begin{figure}
\includegraphics{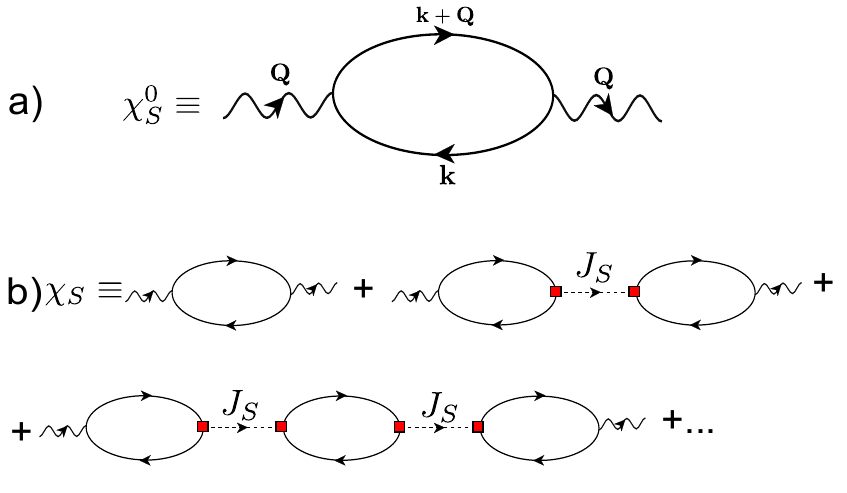}\vspace{-2.5ex}
\caption{\label{chis} (color online) a) The bare polarization bubble we use for the spin mode. For the spin mode, the external momentum equals the AF ordering wave vector $\bf{Q}=(\pi,\pi)$. b) Series of Feynman diagram of the spin mode resulting from the RPA approximation. The spin susceptibility $\chi_{S}$ is the result of this series of diagram which can be evaluated by the formula \ref{eq:3}.}
\end{figure}
\begin{figure}
\includegraphics{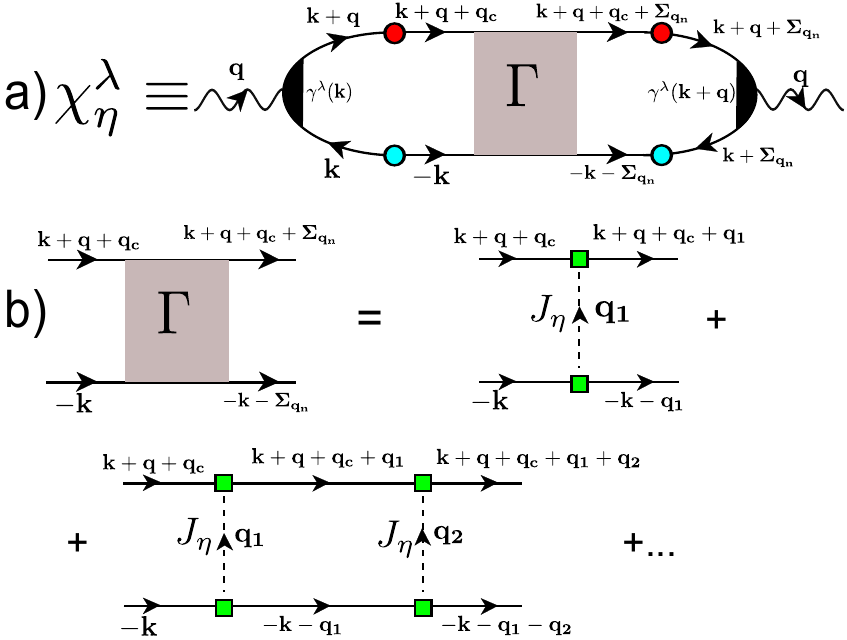}\vspace{-2.5ex}
\caption{\label{chipsi} (color online) a) The Feynman diagram of the $\eta$ collective mode. The red circle represents the CDW quasiparticle Green functions $G^{\Delta_{CDW}}$ of equation (\ref{GCDW}) . The blue circle represents the anomal SC function $G^{\Delta_{SC}}$ of equation (\ref{GSC}). $\gamma_{\bf{k}}^{\lambda}$ is the Raman vertex in the $\lambda$ symmetry. b) $\Gamma$ is the diagram resulting from the Bethe-Salpether series of diagram. $\Sigma_{n}\bf{q_{n}}$ is the summation over the internal momenta $\bf{q_{n}}$. $q_{c}$ is the charge density ware ordering vector. The value of the $\eta$-collective mode susceptibility $\chi_{\eta}^{\lambda}$ is evaluated by the formula (\ref{eq:3}) in the approximation of a frequency and momentum independent $J_{\eta}$.}
\end{figure}
\begin{figure}
\includegraphics{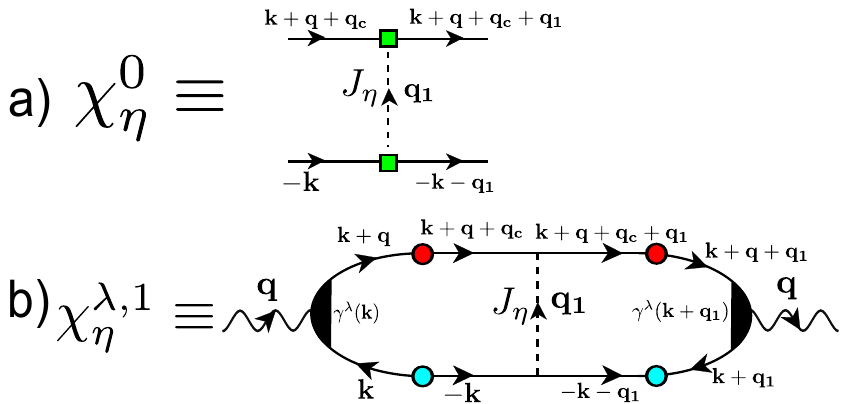}\vspace{-2.5ex}
\caption{\label{chipsid} (color online) a) $\chi_{\eta}^{0}$ is the \textcolor{black}{single segment of the ladder diagrams describing} the $\eta$-mode. This term is evaluated in the formula (\ref{chipsizero}) in the section \ref{chibare}. b) $\chi{_\eta}^{\lambda,1}$ is the first term of the serie of ladder diagram. $\gamma_{\bf{k}}^{\lambda}$ is the Raman vertex in the $\lambda$ symmetry. This term is evaluated in the formula (\ref{chipsi1}) in the section \ref{chif}. Both terms are evaluated in the approximation of a momentum and frequency independent $J_{\eta}$.}
\end{figure}
\textcolor{black}{We rewrite the Hamiltonian (\ref{eq:1}) with the spin operators ($S_{\pm}$ of equation (\ref{eq:2bis})) and the $\eta$-mode operator (see equation (\ref{eq:2})) : $H=\sum_{\mathbf{k},\sigma}\xi_{\mathbf{k}}c_{\mathbf{k}\sigma}^{\dagger}c_{\mathbf{k}\sigma}+J_{S}S_{+}S_{-}+J_{\eta}\eta^{\dagger}\eta$. The magnitude of the interactions are $J_{\eta}=3J-V$ in the $\eta$-mode channel and $J_{S}=2J$ in the spin channel.}
 We compare the spin susceptibility derived from the Random Phase Approximation (RPA) with $\chi_{S}=-i\theta\left(t\right)\left\langle \left[S_{+}\left(t\right),S_{-}\left(0\right)\right]\right\rangle $ as depicted in Fig \ref{chis} and the $\eta$-mode susceptibility derived from Bethe-Salpeter ladder approximation with $\chi_{\eta}=-i\theta\left(t\right)\left\langle \left[\eta\left(t\right),\eta^{\dagger}\left(0\right)\right]\right\rangle $ shown in Fig \ref{chipsi}.
Within the RPA and Bethe Salpether approximation, the full susceptibilities are given by:
\begin{align}
\chi^{\lambda}_{\eta}\left(\omega\right) & =\frac{\chi_{\eta}^{\lambda,1}\left(\omega\right)}{1-J_{\eta}\chi^{0}_{\eta}\left(\omega\right)},\;\;\quad & \chi_{S}\left(\omega\right) & =\frac{\chi_{S}^{0}\left(\omega\right)}{1+J_{S}\chi^{0}_{S}\left(\omega\right)},
\label{eq:3}
\end{align}
where $\chi^{0}_{S}$ is the bare polarization bubble for the spin mode (see Fig. \ref{chis} (a)), $\chi^{0}_{\eta}$ is the \textcolor{black}{single segment of the ladder diagrams describing} the $\eta$-mode (see Fig. \ref{chipsid} (a)) and $\chi^{\lambda,1}_{\eta}$ is the first term of the diagram series for $\eta$-mode (see Fig. \ref{chipsid} (b)) in the $\lambda$ symmetry. Note that the detailed calculation of the bare polarization bubble and diagrams is done in the section \ref{Eval}.
\end{subsubsection}
\end{subsection}
\begin{subsection}{Evaluation of the diagrams}
\label{Eval}
In this section, we focus on the diagram calculation presented in Fig. \ref{chis} (a), for the bare polarization of the spin mode $\chi_{S}^{0}$, in Fig. \ref{chipsid} (a), for \textcolor{black}{the single segment of the ladder diagrams describing} the$\eta$-mode $\chi_{\eta}^{0}$ and in Fig. \ref{chipsid} (b) for the first term of the $\eta$-mode diagram series. The explicit internal and external momenta dependence of each diagram are explicitly presented in  Fig. \ref{chis} (a) and \ref{chipsid}.

\textcolor{black}{In order to include the frequency dependence of the order parameters in the diagram calculation, we develop the order parameters depending on the internal frequency standing inside the summation terms around the external frequency, $\Delta(\epsilon_{int})\approx \Delta(\epsilon_{ext})+(\epsilon_{int}-\epsilon_{ext})\frac{\partial\Delta}{\partial\epsilon_{int}}_{\epsilon_{int}=\epsilon_{ext}}+...$, and we keep only the zeroth order term (with $\epsilon_{int (ext)}$ the internal (external) Matsubara frequency).} This implies that our susceptibility will be under-estimated at high-frequency.

We also assume that the $\eta$-mode interaction $J_{\eta}$ and the spin interaction $J_{S}$ do not depend on the frequency and the momentum. This approximation allow us to decompose the ladder diagram of Fig. \ref{chipsi} in the diagram of Fig. \ref{chipsid}.
The momentum sum is done over the first Brillouin zone with meshes of $400\times400$ after doing the summation over internal Matsubara frequencies at zero temperature $T=0$ . We do the analytical continuation on the external Matsubara frequency writing $i\omega\equiv\omega+i\delta$ in the energy denominators. A small broadening is introduced by the parameter $\delta$ and can be understood as a residual scattering.
\begin{subsubsection}{Calculation of the bare polarization bubbles.}
\label{chibare}
\begin{figure}
\includegraphics{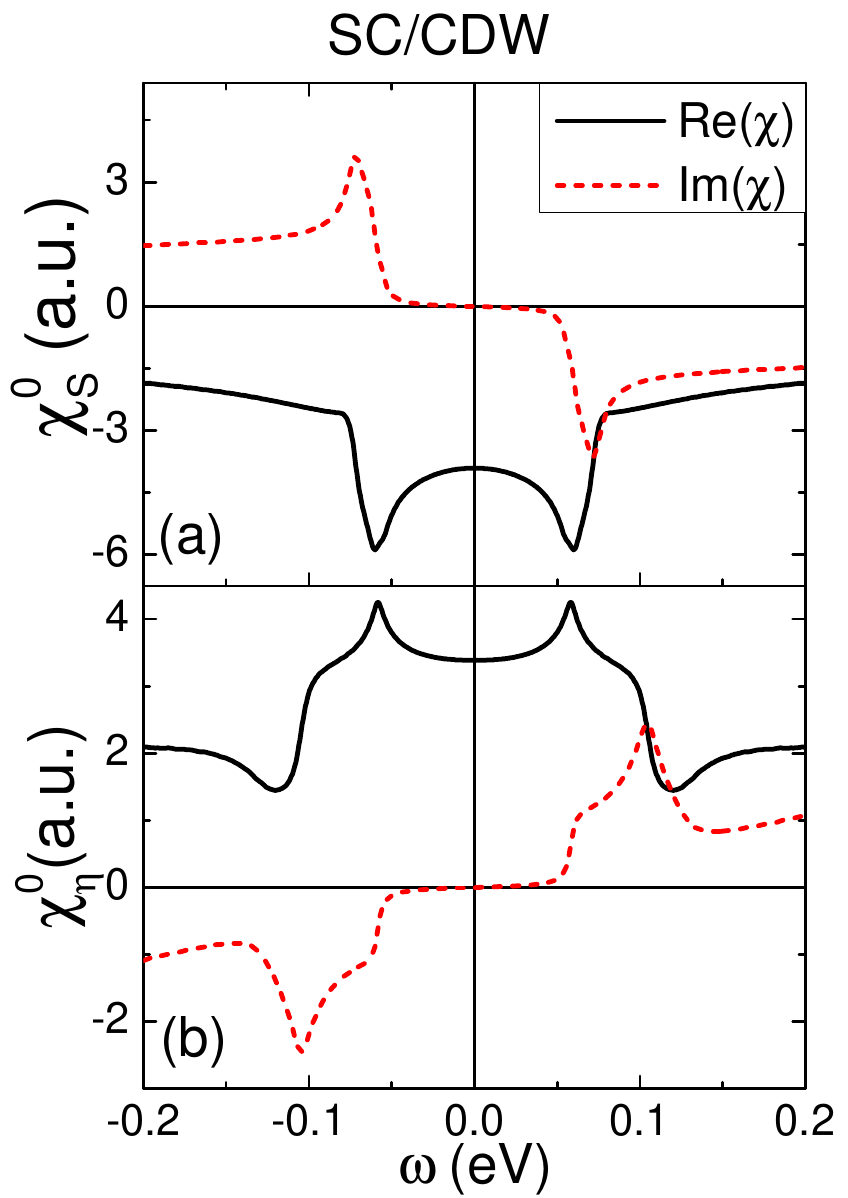}\vspace{-2.5ex}
\caption{\label{baresusceptibilitya}(Color online) Imaginary part (dashed line) and real part (solid line) of (a) the spin bare function $\chi_{S}^{0}$ at $\mathbf{q=Q=(\pi,\pi)}$ and (b) the $\eta$-mode bare function $\chi_{\eta}^{0}$ at $\mathbf{q=}(0,0)$ at $T=0K$ in a d-wave superconductor with 2$\Delta_{SC}^{0}=60meV$ mixed to a d-wave charge order with 2$\Delta^{0}_{CDW}=35meV$ . A broadening $\delta$ of 5meV was employed.}
\end{figure}
\begin{figure}
\includegraphics{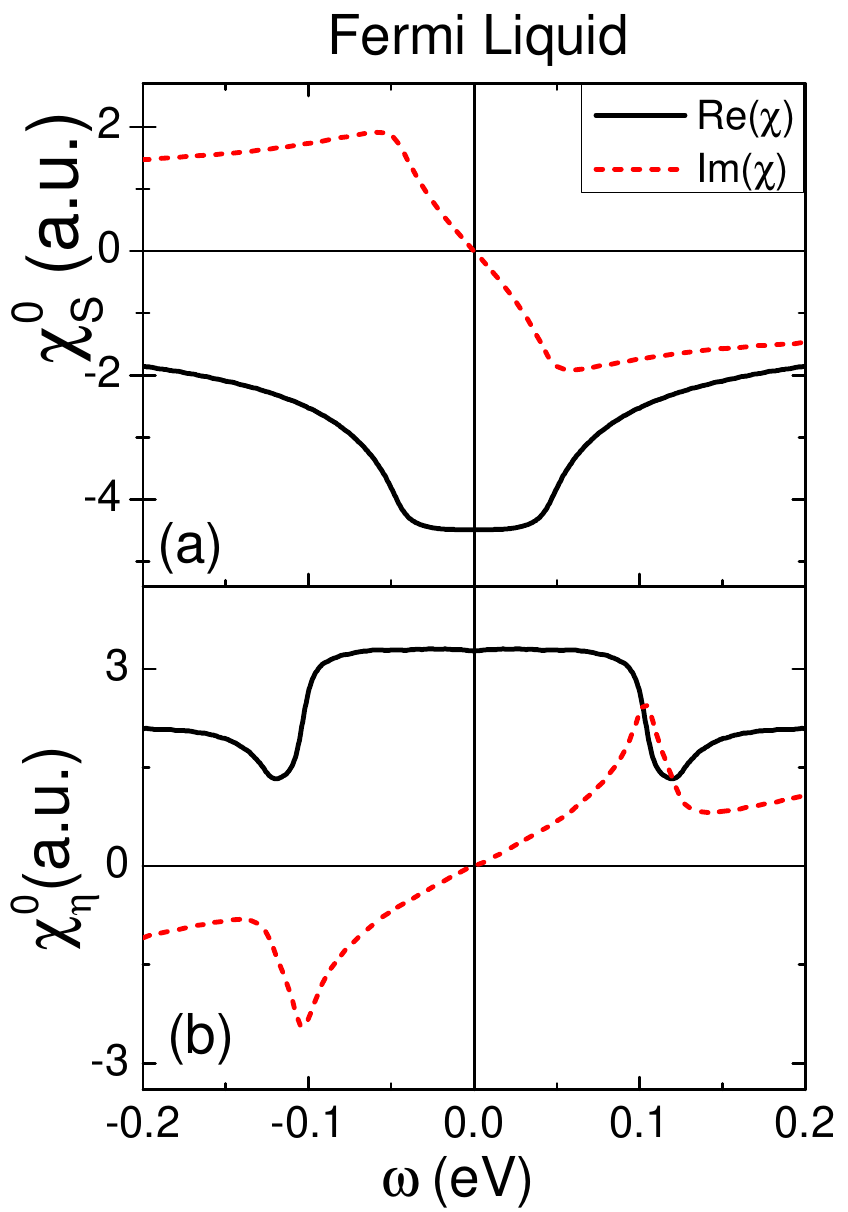}\vspace{-2.5ex}
\caption{\label{baresusceptibilityb}(Color online) Imaginary part (dashed line) and real part (solid line) of (a) the spin bare function $\chi_{S}^{0}$ at $\mathbf{q=Q=(\pi,\pi)}$ and (b) the $\eta$-mode bare function $\chi_{\eta}^{0}$ at $\mathbf{q=}(0,0)$ at $T=0K$ in the Fermi liquid phase. A broadening $\delta$ of 5meV was employed.}
\end{figure}
The bare polarization $\chi^{0}$ is the response function of non local density operator : $\chi^{0}=-i\theta(\tau)\langle\rho_{\mathbf{q}}(\tau)\rho_{\mathbf{q}}(0)\rangle$ with $\rho_{\mathbf{q}}=1/N\sum_{\mathbf{k}}c^{\dagger}_{\mathbf{k}}c_{\mathbf{k+q}}$.
From the figure \ref{chis} (a) and following the academical way to evaluate these diagrams (see for example the references \cite{Schrieffer64} and \cite{AGD65}), we find:
\begin{align}
&\chi_{S}^{0}(\mathbf{q},i\omega)=T\sum_{\epsilon,\mathbf{k}}\text{Tr}\left[\hat{G}(i\epsilon+i\omega,\mathbf{k+q})\cdot\hat{G}(i\epsilon,\mathbf{k})\right],
\end{align}
where $\epsilon (\omega)$ is a fermionic (bosonic) Matsubara frequency, $T$ the temperature and Tr means Trace of the Green function matrix $\hat{G}$ (see expression (\ref{eq:green-1})). $\hat{G}$ is the Green function matrix deduced from the inversion of the matrix (\ref{eq:green-1}).
From the figure \ref{chipsid} a), and assuming that $J_{\eta}$ is frequency and momenta independent, the ladder diagram $\chi_{\eta}^{0}$ can be determined (see \cite{Schrieffer64} and \cite{AGD65}) in the form :
\begin{align}
&\chi_{\eta}^{0}(\mathbf{k,q},\omega,\epsilon)=T\sum_{\omega_{1},\mathbf{q_{1}}}\nonumber\\
&\text{Tr}\left[\hat{G}(i\epsilon+i\omega+i\omega_{1},\mathbf{k+q+q_{1}+q_{c}})\cdot\hat{G}(-i\epsilon-i\omega_{1},\mathbf{-k-q_{1}})\right]
\label{chipsizero}
\end{align}
where $\omega_{1}$ is a bosonic Matsubara frequency. Note that we put $J_{\eta}$ outside of the expression (\ref{chipsizero}). We do the internal Matsubara frequency and momentum summation by doing the change of variable $\bf{\tilde{k}=k+q_{1}}$ and $\tilde{\omega}=\omega_{1}+\epsilon$. Then, $\chi_{\eta}^{0}(\mathbf{k,q},\omega,\epsilon)$ can be simplified as $\chi_{\eta}^{0}(\mathbf{q},i\omega)$ and becomes :
\begin{align}
&\chi_{\eta}^{0}(\mathbf{q},\omega)=T\sum_{\tilde{\omega},\mathbf{\tilde{k}}} \text{Tr}\left[\hat{G}(i\tilde{\omega}+i\omega,\mathbf{\tilde{k}+q+q_{c}})\cdot\hat{G}(-i\tilde{\omega},\mathbf{-\tilde{k}})\right]
\end{align}
The bare polarization in the spin and $\eta$ channels in the SC/CDW coexisting phase are presented in Fig.\ref{baresusceptibilitya} (with 2$\Delta_{SC}^{0}=63\,meV$ and 2$\Delta^{0}_{CDW}=35\,meV$) and in the Fermi liquid phase in Fig. \ref{baresusceptibilityb}. 
\textcolor{black}{The bare polarization in the spin channel $\chi_{S}^{0}$ develops a gap in the SC/CDW phase Fig.\ref{baresusceptibilitya} (a). The presence of a threshold in Im$\chi_{S}^{0}$ at the frequency $\omega\approx60\,meV$ allows the emergence of a resonance of the collective response below the gap. This gap closes in the Fermi liquid phase as seen in Fig.\ref{baresusceptibilityb} (a).}

On the other hand, the $\eta$-mode couples an electron with momentum $\mathbf{k}+\mathbf{q_{c}}$ to a counter propagating electron with momentum $\mathbf{-k}$. Consequently, $\chi_{\eta}^{0}$ has an opposite sign of the one in the spin channel in both Fermi liquid and superconducting state (see Fig.\ref{baresusceptibilitya} (b) and Fig.\ref{baresusceptibilityb} (d)). \textcolor{black}{In the $\eta$ channel $\chi_{\eta}^{0}$ develops a gap in the SC/CDW phase Fig.\ref{baresusceptibilitya} (b). This quasiparticle gap closes in the Fermi liquid phase as seen in Fig.\ref{baresusceptibilityb} (b).} We clearly see a gap with a threshold occurring around $\omega\approx 60\,meV$ which is the value of $2\Delta_{SC}^{0}$. Note that a gapped quasiparticle continuum develops at low frequency in $\chi_{\eta}^{0}$ as in $\chi_{S}^{0}$. The reason is that we evaluated $\chi_{\eta}^{0}$ at $\mathbf{q=0}$ while the CDW order has a non-zero ordering vector $\bf{q_{c}}$. For comparison, $\chi_{S}^{0}$ is evaluated at finite momentum $\mathbf{q=Q}$ which also results in a gapped quasiparticle continuum.
\end{subsubsection}
\begin{subsubsection}{Calculation of the term $\chi_{\eta}^{\lambda,1}$}
\label{chif}
The ladder diagram $\chi_{\eta}^{\lambda,1}$ can be determined (see \cite{Schrieffer64} and \cite{AGD65}) in the form : 
\begin{align}
&\chi_{\eta}^{\lambda,1}(\mathbf{q},\omega)=-T^{2}\sum_{\epsilon,\omega_{1},\mathbf{k,q_{1}}}J_{\eta}(\omega_{1},\mathbf{q_{1}})\gamma^{\lambda}_{\bf{k}}\gamma^{\lambda}_{\bf{k+q+q_{1}}} \times \nonumber\\
&\left[G^{\Delta_{CDW}}(i\epsilon+i\omega,\textbf{k+q})\cdot G^{\Delta_{SC}}(i\epsilon,\textbf{k}) \times \right. \nonumber\\
& \left. G^{\Delta_{CDW}}(i\epsilon+i\omega+i\omega_{1},\mathbf{k+q+q_{1}+q_{c}}) \times \right. \nonumber\\
& \left. G^{\Delta_{SC}}(-i\epsilon-i\omega_{1},\mathbf{-k-q_{1}})\right]
\label{chipsi1}
\end{align}
Considering our initial approximation, we can factorize by the interaction $J_{\eta}$ and we determine, from the Green function matrix (\ref{eq:green-1}), that $G^{\Delta_{SC}}(-i\epsilon-i\omega_{1},\mathbf{-k-q_{1}})=G^{\Delta_{SC},\star}(i\epsilon+i\omega_{1},\mathbf{k+q_{1}})$ and $G^{\Delta_{CDW}}(i\epsilon+i\omega+i\omega_{1},\mathbf{k+q+q_{1}+q_{c}})=G^{\Delta_{CDW},\star}(i\epsilon+i\omega+i\omega_{1},\mathbf{k+q+q_{1}})$. We do the internal Matsubara frequency and momentum summation by doing the change of variable $\bf{\tilde{k}=k+q_{1}}$ and $\tilde{\omega}=\omega_{1}+\epsilon$.

\textcolor{black}{This susceptibility is proportional to the product of both SC and CDW order parameters $\Delta_{CDW}.\Delta_{SC}$. Consequently, the $\eta$-mode susceptibility $\chi^{\lambda,1}_{\eta}$ only exists in a coexisting SC/CDW phase. A careful study of the symmetry of the $\eta$-mode response is done in the following section.}

\end{subsubsection}
\end{subsection}
\begin{subsection}{The full Raman response}
\label{Raman}
\begin{figure}
\includegraphics{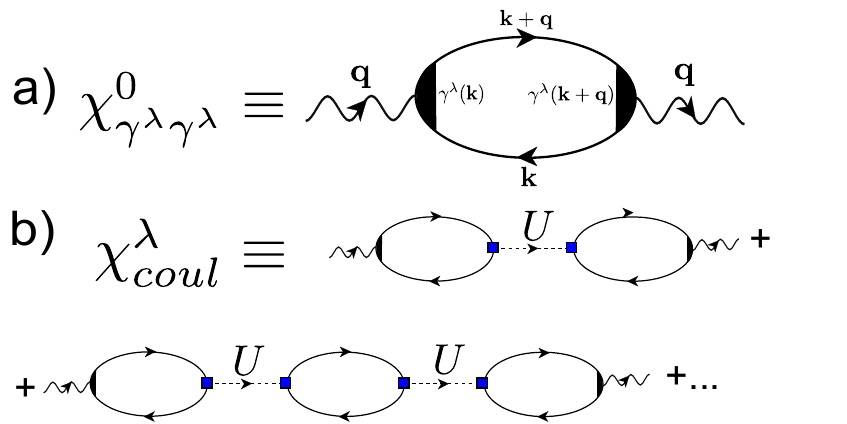}\vspace{-2.5ex}
\caption{\label{chiraman}(Color online) a) The Raman bare polarization bubble with $\gamma^{\lambda}$ the Raman vertex in the $\lambda$ symmetry. b) $\chi^{\lambda}_{coul}$ is the Coulomb screening term resulting from the RPA series of diagram. Note that the dark triangle in the diagram denotes the presence of the Raman vertex.}
\end{figure}
We now argue that the $\eta$-collective mode is seen in the $A_{1g}$ Raman channel only.
The study of the Raman susceptibility requires a careful examination of the symmetries of the system. These symmetries are taken into account by considering vertices in the photon-matter interaction different from unity (see Fig. \ref{chipsi} and \ref{chiraman}). Three symmetries are typically considered for the Raman vertices in the cuprate superconductors which write within the effective mass approximation:
\begin{align}
&\gamma^{A_{1g}}=\frac{1}{2}\left[\frac{\partial^{2}\xi_{k}}{\partial^{2}k_{x}}+\frac{\partial^{2}\xi_{k}}{\partial^{2}k_{y}}\right];
\gamma^{B_{1g}}=\frac{1}{2}\left[\frac{\partial^{2}\xi_{k}}{\partial^{2}k_{x}}-\frac{\partial^{2}\xi_{k}}{\partial^{2}k_{y}}\right]\nonumber\\
&\gamma^{B_{2g}}=\frac{1}{2}\left[\frac{\partial^{2}\xi_{k}}{\partial k_{x} \partial k_{y}}+\frac{\partial^{2}\xi_{k}}{\partial k_{y} \partial k_{x}}\right]
\label{eq:4}
\end{align}
where $A_{1g}$ probes the whole Brillouin zone, $B_{1g}$ the antinodal zone and $B_{2g}$ the nodal zone. Here we assume that there is no response in the $A_{2g}$ symmetry \cite{Devereaux-RMP} ($\gamma^{A_{2g}}=0$). In presence of the $\eta$ collective mode, the full Raman susceptibility writes in the $\lambda$ symmetry:
\begin{align}
\chi_{Raman}^{\lambda}=\chi^{0}_{\gamma^{\lambda} \gamma^{\lambda}}+\chi^{\lambda}_{coul}+\chi^{\lambda}_{\eta}
\label{eq:5}
\end{align}
where $\chi^{0}_{\gamma^{\lambda} \gamma^{\lambda}}$ is the bare Raman response (Fig. \ref{chiraman} a)), $\chi^{\lambda}_{coul}$ the Coulomb screening (Fig. \ref{chiraman} b)) and $\chi^{\lambda}_{\eta}$ the $\eta$-mode response (Fig. \ref{chipsi} a)).
 
\textcolor{black}{The value of $\chi^{\lambda}_{\eta}$ depends on the momentum dependence of the product $\gamma^{\lambda}_{\bf{k}}G^{\Delta_{CDW}}\cdot G^{\Delta_{SC}}$ .
The product of the two Green functions $G^{\Delta_{CDW}}\cdot G^{\Delta_{SC}}$ is proportional to the product $\Delta_{SC}.\Delta_{CDW}$ which is proportional to the square of the $d$-wave factor.
The sum over the internal momenta in the FBZ of the square of the $d$-wave factor is finite. The sum over the internal momenta in the FBZ of the Raman vertex $\gamma^{B_{1g}}_{\bf{k}}$ and $\gamma^{B_{2g}}_{\bf{k}}$ vanishes ($\sum_{k}\gamma^{B_{1g}}_{k}=\sum_{k}\gamma^{B_{2g}}_{k}=0$) while it is finite for the Raman vertex $\gamma^{A_{1g}}_{\bf{k}}$ ($\sum_{k}\gamma^{A_{1g}}_{k}\neq0$). 
The sum over the internal momenta of the product $\gamma^{\lambda}_{\bf{k}}G^{\Delta_{CDW}}\cdot G^{\Delta_{SC}}$ therefore vanishes in the $B_{1g}$ and $B_{2g}$ symmetry while it is non-zero in the $A_{1g}$ symmetry. Consequently, the $\eta$-mode response only exists in $A_{1g}$ symmetry.}

In addition long-range Coulomb interaction $U\sim1/q^{2}$ plays an important role in screening the Raman in the $A_{1g}$ channel \cite{Devereaux-RMP}. Doing the summation over all the coulomb diagram (Fig. \ref{chiraman} b), one can find easily:
\begin{align}
\chi^{\lambda}_{coul}=\frac{U\chi^{0}_{\gamma^{\lambda} 1}\chi^{0}_{1 \gamma^{\lambda}}}{1-U\chi^{0}_{11}}
\label{eq:chicoul}
\end{align}
where $\chi^{0}_{\gamma^{\lambda} 1}$ is the Raman susceptibility with one of the vertex put to unitiy and $\chi^{0}_{11}$ the bare polarization bubble.
In the limit $q\rightarrow0$, the "Coulomb screened" susceptibility simplifies as $\chi^{\lambda}_{coul}=-\chi^{0}_{\gamma^{\lambda} 1}\chi^{0}_{1 \gamma^{\lambda}}/\chi^{0}_{11}$.
In the $A_{1g}$ symmetry, the contribution of $\chi^{A_{1g}}_{coul}$ cannot be neglected \cite{cardona97}. Its contribution will screen partly the bare Raman susceptibility $\chi^{0}_{\gamma^{A_{1g}} \gamma^{A_{1g}}}$. Consequently, the Raman response in the $A_{1g}$ symmetry writes $\chi_{Raman}^{A_{1g}}=\chi^{0}_{\gamma^{A_{1g}} \gamma^{A_{1g}}}+\chi^{A_{1g}}_{coul}+\chi^{A_{1g}}_{\eta}$. Around optimal doping, we will see that the contribution of the $\eta$-mode (last term) is of the same order of magnitude as the bare screened $A_{1g}$ Raman response (first two terms). By increasing the doping, because of the weakening of the CDW order we expect a decrease of the $\eta$-mode intensity and the A$_{1g}$ spectra will be dominated by the bare screened Raman response.

The Coulomb screening can be neglected in the $B_{1g}$ and $B_{2g}$ channels for symmetry reasons \cite{cardona97}. Consequently, the full Raman susceptibility is simply given by the unscreened response: $\chi^{\lambda}_{Raman} \approx \chi^{0}_{\gamma^{\lambda} \gamma^{\lambda}}$ with $\lambda = B_{1g}(B_{2g})$.
\end{subsection} 
\end{section}
\begin{section}{Results and Discussion}
\label{results}
\textcolor{black}{The theoretical spectra are calculated without solving the self-consistent equation derived from the spin-fermion model. As a consequence, we need to adjust the amplitude of the order parameters as well as the value of $J$ and $V$ to reproduce the experimental data.
The parameters values are adjusted in the following way : the value of $\Delta_ {SC}^{0}$ is chosen to reproduce the Raman peak frequency in the $B_{1g}$ channel. The value of $\Delta_ {CDW}^{0}$ is chosen to adjust the intensity of the resonance in the $A_{1g}$ channel. Note that the value of $\Delta_{CDW}^{0}$ does not influence the frequency of the $\eta$-mode.
We adjust the value of $J$ to fit the neutron resonance frequency and the value of $V$ to fit frequency of the Raman resonance in the $A_{1g}$ symmetry.
A full self-consistent determination of the CDW and SC order parameter amplitude is not done here as our goal is to identify the main theoretical features necessary to explain the Raman resonance in the $A_{1g}$ symmetry.}
\begin{subsection}{Results}
\begin{figure}
\includegraphics{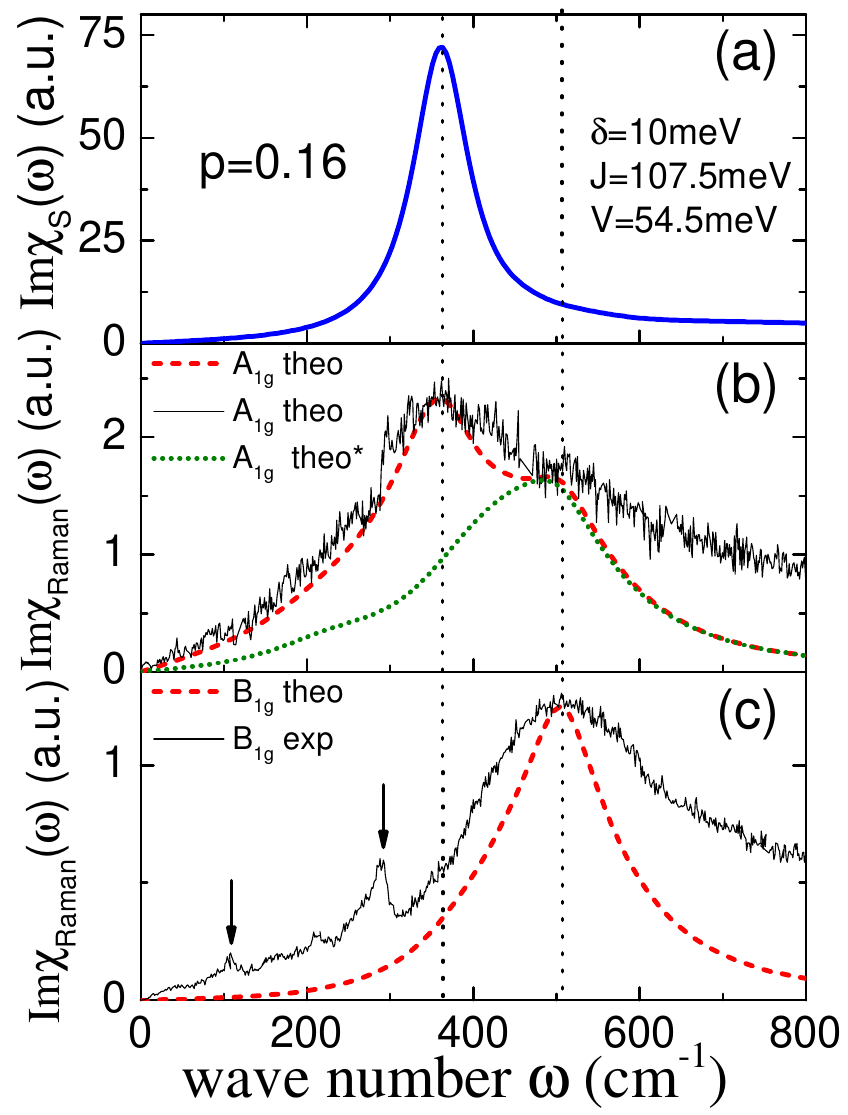}
\caption{\label{SupA} (a) Calculated spin response. (b) Experimental  (solid line) and calculated (dashed line) Raman response in the $A_{1g}$ symmetry and (c) in the $B_{1g}$ symmetry at optimal doping $p=0.16$ ($\delta=10\,meV$, 2$\Delta_{SC}^{0}=63 \,meV,2\Delta^{0}_{CDW}=45 \,meV$). In (b), the dotted line presents the calculated bare screened Raman response without the $\eta$ mode contribution, $\chi^{A_{1g}\star}=\chi^{0}_{\gamma^{A_{1g}} \gamma^{A_{1g}}}+\chi^{A_{1g}}_{coul}$. Phonon lines have been subtracted out of the experimental Raman spectra for clarity in (b) but not in (c). The arrows in (c) indicate the location of the phonon lines superimposed on the electronic background.}
\end{figure}
\begin{figure}
\includegraphics{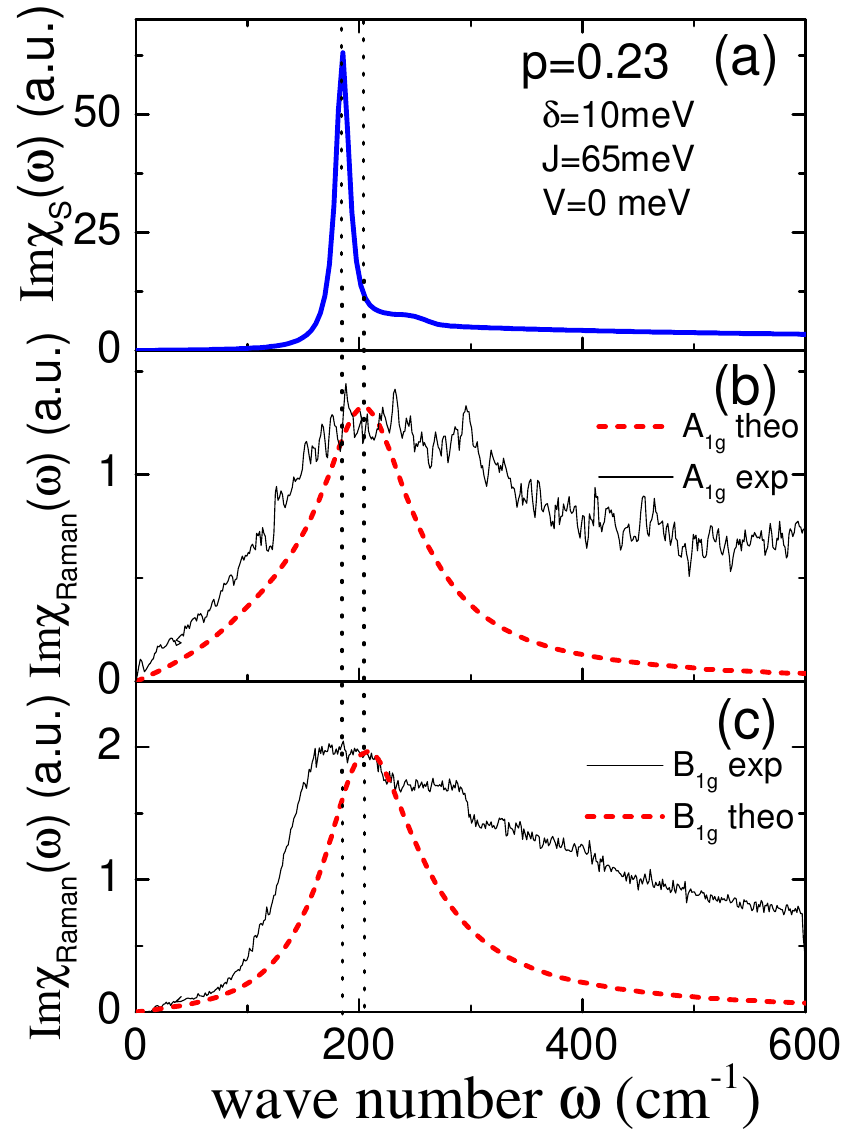}
\caption{\label{SupB} (a) Calculated spin response. (b) Experimental (solid line) and calculated (dashed line) Raman response in the $A_{1g}$ symmetry and (c) in the $B_{1g}$ symmetry for the overdoped regime $p=0.23$ ($\delta=10\,meV$, 2$\Delta_{SC}^{0}=25 \,meV,2\Delta^{0}_{CDW}=0 \,meV$). Phonon lines have been subtracted out of the experimental A$_{1g}$ Raman spectra in (b) for clarity.}
\end{figure}

The imaginary parts of the neutron spin susceptibility (spin response) and the full Raman susceptibility in the $A_{1g}$ symmetry (Raman response) are presented in (a) and (b) panels of Fig.\ref{SupA} for optimal doping (p=0.16) with $2\Delta_{SC}^{0}=63\,meV$ and $2\Delta_{CDW}^{0}=45\,meV$. The theoretical curves are obtained for $J=107.5\,meV$ and $V=54.5\,meV$. \textcolor{black}{To the best of our knowledge, the amplitude of the CDW order parameter $\Delta^{0}_{CDW}$ has not been evaluated yet in the experiments. However, in our effective model, the value of $\Delta^{0}_{CDW}$ is reasonable compared with the SC order parameter amplitude.}

For this set of parameters, the calculated spin response exhibits a sharp peak at $\omega=364cm^{-1}$ at the same energy than the energy of the calculated $A_{1g}$ Raman resonance. Note that the frequency dependence of the CDW order $\Delta_{CDW}$ cuts the SC coherence peak at $2\Delta^{0}_{SC}$ $\sim$ $(510 \,cm^{-1}(63\,meV)$ in the $A_{1g}$ channel but leaves the $B_{1g}$ channel unaffected (see panel (c)). Fig.\ref{SupA} (b) and (c) show a qualitative agreement between the peak energies in the $A_{1g}$ and $B_{1g}$ calculated Raman spectra (red curves) and the experimental Raman spectra obtained from Bi$2212$ single crystals (black curves) \cite{sacuto15}. \textcolor{black}{We note that in our approximation, the calculated susceptibility vanishes in the metallic phase. This implies that we cannot reproduce the susceptibility at high energy for $\omega>$2$\Delta^0_{SC}$. More realistic electron self-energies are needed in order to reproduce the spectra at high energy.}

In the over-doped regime (p=0.23), the value of $J=65\,meV$ decreases and the value of $V$ is the same than at optimal doping. Experimentally above optimal doping, the frequency of the INS resonance at $\bf{Q}$ decreases with overdoping and scales with $T_{c}$ \cite{Sidis2001}. \textcolor{black}{In absence of experimental data, we choose to fit the neutron scattering peak by assuming that the scaling between the neutron resonance frequency and $T_{c}$ postulated in \cite{Sidis2001} is still valid at hole doping $p=0.23$. At this doping, the INS resonance frequency should be at $187\,cm^{-1}$ with $T_{c}=50K$.} \textcolor{black}{Note that persistent spin excitation have been observed in a huge range of hole doping by resonant inelastic X-Ray scattering experiments \cite{LeTacon11,Dean112013} which favours the existence of neutron resonance at $\bf{Q}$ vector at high doping.} The maximal value of $\Delta_{SC}$ and $\Delta_{CDW}$ weakens and we take into account this decreasing by putting $2\Delta_{SC}^{0}=25\,meV$ and $2\Delta_{CDW}^{0}=0\,meV$. The decrease of $\Delta_{CDW}^0$ is due to the larger distance to the AF QCP at this doping as well as the proximity of a change of Fermi surface topology \cite{sacuto15,kaminski2006}. The vanishing of the CDW order in overdoped regime has been observed by STM \cite{Fujita14}.
\par
For these parameters the calculated spin and A$_{1g}$ Raman responses exhibit  peaks at different energies (187$\,cm^{-1}$ and 205\,cm$^{-1}$ respectively) ((see Fig.\ref{SupB} (a) and (b)). The perfect matching between Neutron and Raman peak energy is thus only verified close to optimal doping level.
The theoretical Raman response capture well the energies of both the $B_{1g}$ SC coherence peak $2\Delta_{SC}^{0}$ and the $A_{1g}$ peak detected in the experimental spectra shown in Fig.\ref{SupB} (b) and (c).

The $\eta$-mode also exists in the overdoped regime but the decreasing of $\Delta_{SC}$ and $\Delta_{CDW}$ completely weakens its intensity. The $2\Delta$ SC coherence peak, which is not detectable in the optimal doped regime (see Fig.\ref{SupA} (b)), increases its intensity with overdoping and become detectable in the A$_{1g}$ channel around $p=0.22$. Beyond this doping, the CDW order parameter amplitude vanishes and the $\eta$-mode contribution to the A$_{1g}$ Raman spectrum vanishes. At this doping levels the A$_{1g}$ spectra are entirely given by the bare screened Raman response which peaks at a different energy than the neutron resonance.
To summarize, the physical mechanism behind the $A_{1g}$ resonance peak is different below $p=0.22$ than above. Below this doping, the $A_{1g}$ resonance  originates the $\eta$-mode. Above this doping, the $\eta$-mode vanishes with the CDW order and the peak in the $A_{1g}$ channel is a $2\Delta$ SC coherent peak located at the same energy as the one in the $B_{1g}$ channel. This is qualitatively consistent with Raman experiments, more neutron data in the overdoped regime would be desirable to confirm this picture. \textcolor{black}{Note that in the Fig.\ref{SupA} and \ref{SupB}, we adjust the units to superimpose the experimental and calculated data. Consequently, we cannot comment on the spectral weight of the results.}

\textcolor{black}{ Before proceeding with the discussion,  we briefly comment on the choice of parameters.  In the AF state, ab-initio calculations determine a superexchange coupling $J_{theo}^{AF}=125\,meV$ \cite{Calzado2000,Munoz2002,Appel1987}. An experimental value of superexchange coupling $J_{AF}^{exp}$ has also been deduced from Raman scattering experiments in Bi2212 compounds and evaluated at $J\approx 125meV$ \cite{sugai89,Blumberg97} in the AF phase.
With hole doping, it is reasonable to assume that the superexchange coupling decreases $J<J_{AF}$ \cite{Blumberg97} because of the screening caused by the presence of holes. The values of $J$ we deduce from the fits (see Fig.\ref{SupA} and Fig.\ref{SupB}) are comparable with the calculated and the experimental values, and are coherent with a decrease of the superexchange coupling with the hole doping.}  
\textcolor{black}{The value of $V$ is weaker than the superexchange $J$. To the best of our knowledge, this value has not been evaluated yet. The non zero value of the $V$ parameter implies that the charge channel is important to explain the $A_{1g}$ Raman resonance.}
\textcolor{black}{The frequency of the $A_{1g}$ Raman resonance depends on the values of $V$ and $J$ since it depends on $J_{\eta}=3J-V$. Note that the $A_{1g}$ Raman frequency is less sensitive to the value of $V$ than to the value of $J$.} \textcolor{black}{Note that our effective model does not involve the strong correlation existing in the system. The value of $J$ and $V$ reported here are then effective parameters.}
\end{subsection}
\begin{subsection}{Discussion}
\textcolor{black}{In this part, we discuss our theoretical results and the main experimental features that characterize the Raman spectra in the $A_{1g}$ and $B_{1g}$ channels in the cuprates. }

\textcolor{black}{1) \textcolor{black}{The Raman experiments in the $B_{1g}$ symmetry well established since a long time the presence of the superconducting coherent peak at the frequency of twice maximum of the $d$-wave SC gap $2 \Delta^{0}_{SC}$ \cite{Cooper88,Staufer92,Sacuto97,Gallais04,LeTacon05b}}. Note that the value of $\Delta^{0}_{SC}$ can be determined by STM experiments \cite{Fujita14}. \textcolor{black}{Our theoretical model reproduces well  this superconducting coherent peak for a large range of doping (see Fig \ref{SupA} and \ref{SupB}) \cite{Sacuto2015TS}.}}

\textcolor{black}{2) In Raman experiments at optimal doping, the Raman A$_{1g}$ resonance appears at a frequency below the superconducting coherent peak frequency $2 \Delta^{0}_{SC}$ \cite{Cooper88,Staufer92,Sacuto97,Gallais04,LeTacon05b} and at an energy very close to the INS resonance frequency at $\bf{Q}=(\pi,\pi)$ \cite{SacutoSidis02,LeTacon06b}. Our model explain this resonance by a $\eta$-collective mode which rotates the SC ground state into a CDW state. The energy of this resonance depends on the interaction amplitude $J$ and $V$ as shown in Fig. \ref{SupA} (a) and (b). \textcolor{black}{Althought our model can reproduce the nearly identical energies of the A$_1g$ Raman resonance and the neutron resonance, the equality is obtained by adjusting the parameter V. Nevertheless our model is able to reproduce this equality for a reasonable set of parameters}}

\textcolor{black}{3) In the overdoped regime, the Raman A$_{1g}$ resonance is experimentally observed at an energy very close to the superconducting coherent peak frequency $2 \Delta^{0}_{SC}$. In our calculations, the $\eta$ collective mode vanishes in this regime because of the disappearance of the CDW order. Consequently, only the bare screened susceptibility contributes to the Raman response and the $A_{1g}$ resonance becomes a superconducting coherence peak as shown in Fig. \ref{SupB} (a) and (b). }

\textcolor{black}{4) \textcolor{black}{In the underdoped regime, the transition temperature of the CDW order is greater than the superconducting critical temperature. Hence we can consider that the CDW order parameter amplitude becomes larger than the SC one $\Delta_{CDW}^{0}>\Delta_{SC}^{0}$.} In our theory this regime implies a decreasing of the $\eta$-mode resonance intensity which will become undetectable at low doping. Existing Raman data seems to be consistent with a disappearance of the A$_{1g}$ in the underdoped regime \cite{Sacuto2015TS}. The description of the underdoped regime however requires a description of the pseudogap phase that we did not take into account in the present work. Deeper considerations about the Raman scattering spectra in the underdoped regime will be considered in details in a forthcoming work.}

\textcolor{black}{5) To fit the Raman resonance in the $A_{1g}$ symmetry, we propose an explanation based on the coexistence between two orders with a $d$-wave symmetry. This $d$-wave symmetry remains crucial to explain the resonance in the  $A_{1g}$ symmetry only. Moreover, the $d$-wave form factor of both SC and CDW orders has been observed by STM \cite{Hamidian15}. \textcolor{black}{To explain the small amplitude of the $2\Delta$ superconducting coherent peak in the $A_{1g}$ symmetry, it is necessary to include a frequency dependence of the order parameters that can be predicted by the spin-fermion model. This frequency dependence suppresses the superconducting coherent peak from the $\eta$-mode response. In the bare Raman response, this superconducting coherent peak is strongly screened by the long-range Coulombian effects. This screening allows the observation of the $\eta$-mode in this channel.}}

\textcolor{black}{6) Experimentally, the $A_{1g}$ Raman resonance has been observed in the SC state and disappear at $T_{c}$ \cite{Cooper88,Staufer92,Sacuto97,Gallais04,LeTacon05b}. The $\eta$-mode calculated here depends on the magnitude of the CDW and SC order parameters and then we expect to see its disappearance for temperature below $T_{c}$ in the overdoped part of the phase diagram (where $\Delta_{SC}^{0}>\Delta^{0}_{CDW}$). This is a limitation of the approach developed here. However, we argue that the $\eta$-mode exists between SC order and all type of charge orders. In our simplified model, we neglect the contributions that could come from the PG phase. In the framework of the interpretation developed in Ref. \cite{Kloss15a}, where the PG phase is interpreted as a charge order, one could observe the emergence of such $\eta$ resonance at higher temperature than the CDW critical temperature.}

\textcolor{black}{}
\end{subsection}
\end{section}
\begin{section}{Conclusion}
\label{conclusion}
To conclude, we propose a coherent scenario to explain the resonance peak in the $A_{1g}$ symmetry seen in Raman scattering. This scenario is based on the existence of coexisting $d$-wave SC and $d$-wave CDW orders in the SC phase of the Cuprates. \textcolor{black}{ The ubiquitous proximity of the CDW and SC states in the underdoped phase of the cuprates generates a collective ``$\eta$-mode'' allowing rotations between those two states.} The $\eta$ collective mode results from the coupling between the  $d$-wave SC and the  $d$-wave CDW state. This mode produces a response in Raman scattering spectroscopy solely in the $A_{1g}$ symmetry and matches the spin triplet resonance at $\mathbf{q=Q=}(\pi,\pi)$ at optimal doping. Since the $\eta$-mode is a charge $\pm2$ spin zero spin singlet excitation other probes like Electron Energy Loss Spectroscopy and Resonant X-Ray techniques are also likely to show the resonance. 
\end{section}

\begin{acknowledgments}
\textcolor{black}{We are grateful for the hospitality of the KITP, Santa Barbara, where discussions at the origin of this work took place. Discussions with M. Norman, I. Paul, A.\ Chubukov, H.\ Alloul, Y.\ Sidis, P.\ Bourges and A.\ Ferraz are also acknowledged. This work was supported by LabEx PALM (ANR-10-LABX-0039- PALM), of the ANR project UNESCOS ANR-14-CE05-0007, as well as the grant Ph743-12 of the COFECUB which enabled frequent visits to the IIP, Natal. X.M. and T.K. also acknowledge the support of CAPES and funding from the IIP.}
\end{acknowledgments}
\begin{appendix}
\clearpage{}
\begin{section}{Supplementary Informations about the spin-fermion model}
\label{Frequency}
In this section, we present the calculation of the effective Lagrangian derived from the initial spin-fermion Lagrangian. The goal of this section is to give more physical details for the completeness of the paper.
\begin{subsection}{The spin-fermion model and the effective boson propagator}
 The starting point of the theoretical model is the spin-fermion model \cite{Metlitski10b,Efetov13}. Note that the models presented in \cite{Metlitski10b,Efetov13} simplify the Fermi surface (FS) in Eight Hot Spots (EHS) (the hot spots are points of the Fermi surface separated by the vector $\textbf{Q} = (\pi,\pi)$). In the following, we develop the model introduced in \cite{Efetov13} in the whole Brillouin zone (BZ).

 In the model \cite{Efetov13}, the physics of cuprates results from the coupling of the fermions with a bosonic spin mode which generates at the Antiferromagnetic(AF)/PseudoGap(PG)  Quantum Critical Point (QCP) a pseudogap in the fermionic dispersion. This new state, different from the AF one, is a superposition of a $d$-wave superconducting (SC) state and a Charge Density Wave (CDW) order. This bosonic spin mode (also called paramagnons)  survives out of the AF phase and become the new glue between electrons that leads to the PG and superconducting (SC) phase. To describe the interaction between this bosonic spin mode and fermions, we start from the Lagrangian $L=L_{c}+L_{\phi}$, where
\begin{align}
&L_{c}=c^{*}(\partial_{\tau}+\xi_{k}+\lambda\phi\sigma)c\\
&L_{\phi}=\frac{1}{2}\phi (\frac{\omega^{2}}{v_{s}^{2}}+(\textbf{q}-\textbf{Q})^{2}+m) \phi+\frac{u}{2}(\phi^{2})^{2}
\end{align}
The Lagrangian $L$ describes the coupling of the electron (represented by the fermionic field $c$) to spin waves (represented by the bosonic field $\phi$). The dynamic of the fermions and their coupling with the bosonic mode is described in the Lagrangian $L_{c}$. The strength of this coupling is tune by the parameter $\lambda$ which couple $c$ and $\phi$. The fermionic dispersion is described by $\xi_{k}=\epsilon_{k}-\mu$ where $\epsilon_{k}$ is the fermionic dispersion and $\mu$ the chemical potential and afford us to describes the Fermi surface of the system. 

The bosonic spin mode physics is described in the Lagrangian $L_{\phi}$. $v_{s}$ is the velocity of the spin waves, $\omega$ the bosonic Matsubara frequency, the paramagnon mass $m$ that describes the distance to the QCP ($m>0$ in the metallic side and vanishes at the QCP) and $\textbf{Q} = (\pi,\pi)$ the ordering wave vector in the AF phase.

The spin wave propagator $D_{q}^{-1}=\left( \frac{\omega^{2}}{v_{s}^{2}}+(\textbf{q}-\textbf{Q})^{2}+m\right)$ is renormalized by the particle-hole bubble (see details in Ref. \cite{Efetov13}). This renormalization leads to consider the effective spin wave propagator $D_{eff}$ that writes $D_{eff,k}^{-1}=(\gamma|\omega|+|\textbf{k}|^2+m)$ where $\gamma$ a phenomenological coupling constant and $\textbf{k}$ has formally been shifted by $\textbf{Q}$.

In the following, we simplify the notation as $k\equiv (i\omega,\bf{k})$, where $i\omega$ are fermionic Matsubara frequencies. To integrate out the bosonic degrees of freedom, we neglect the spinwaves interaction $(u=0)$. The partition function then writes
\begin{align}
&Z =\int \text{d}[\Psi]\text{exp}(-S_{0}-S_{int})\\
&S_{0}=\sum_{k,\sigma}\Psi^{\dagger}_{k,\sigma}G^{-1}_{0,k,\sigma}\Psi_{k,\sigma},\\
&S_{int}=-\sum_{k,k',q,\sigma}J_{q}c^{\dagger}_{k,\sigma}c_{k+Q+q,\bar{\sigma}}c^{\dagger}_{k',\bar{\sigma}}c^{\dagger}_{k'-q-Q,\sigma}.
\end{align}
where the bare propagator is 
\begin{align}
\hat{G}^{-1}_{0k}=\left(\begin{array}{cccc}
 i\omega-\xi_{k}&0&0&0\\
 0&i\omega+\xi_{-k+q_{c}}&0&0\\
 0&0&i\omega-\xi_{k-q_{c}}&0\\
 0&0&0&i\omega+\xi_{-k}
\end{array}\right)
\end{align}
and the spinor field $\Psi_{k}=(c_{k,\sigma},c^{\dagger}_{-k+q_{c},\bar{\sigma}}c_{k-q_{c},\sigma}c^{\dagger}_{-k,\bar{\sigma}})$. Moreover, $J_{q}^{-1}=4D_{q}^{-1}/3\lambda^{2}$, $\sigma=\{\uparrow,\downarrow\}$ labels the spin and $\bf{q_{c}}$ stands for the charge density wave ordering vector.
\end{subsection}
\begin{subsection}{The mean field decoupling and the effective Lagrangian}
After the integration on the bosonic degrees of freedom, we decouple the two-body action $S_{int}$ using mean field decoupling. Thus, we decouple the interaction Lagrangian in the SC channel and in the charge channel by introducing the two order parameters $\Delta_{k}$ and  $\Delta_{CDW,k}$.
$\Delta_{k}$ is the superconducting order parameter which describes the $d$-wave SC order. $\Delta_{CDW,k}$ is the CDW order parameter which describes the $d$-wave charge order with ordering vector $\bf{q_{c}}$. Note that in the reference \cite{Efetov13}, these two order parameters are related to the SU(2) symmetry which is verified at the hot-spot. The SU(2) symmetry between SC and CDW order exists if their energy level is the same (the energy splitting $\Delta_{k}-\Delta_{CDW,k}$ vanishes). In the whole Brillouin zone, the SU(2) symmetry is verified around the hot-spots (at low paramagnon mass) and at the zone edge (at high paramagnon mass) \cite{Kloss151}.

Applying the Hubbard-Stratanovitch transformation to the interaction action $S_{int}$, the partition function becomes
\begin{align}
&Z =\frac{\int d[\Psi]d[\Delta,\Delta_{CDW}]exp(-S_{0}-S_{int,eff})}{\int d[\Delta,\Delta_{CDW}]exp(-S_{q})}\\
&S_{q}=-\sum_{k,q,\sigma}[J_{q}^{-1}\Delta_{CDW,k}^{\dagger}\Delta_{CDW,\bar{k}+q}+J_{q}^{-1}\Delta_{k}^{\dagger}\Delta_{\bar{k}+q}]\\
&S_{int,eff}=S_{q}-\sum_{k,\sigma}\Psi^{\dagger}_{k,\sigma}\hat{M}_{k}\Psi_{k,\sigma},
\end{align}
with $\bar{k}=k+Q$ and the matrix $\hat{M}_{k}$ is
\begin{align}
\hat{M}_{k}=\left(\begin{array}{cc}
 & \hat{m}_{k}\\
\hat{m}^{\dagger}_{k} & 
\end{array}\right),
\hat{m}_{k}=\left(\begin{array}{cc}
 -\Delta_{CDW,k}& -\Delta_{k}\\
-\Delta^{\dagger}_{k+q_{c}} &\Delta_{CDW,-k} 
\end{array}\right).
\end{align}
Finally, the effective Lagrangian we will use in the following will be defined as $\hat{G}^{-1}=\hat{G}^{-1}_{0k}-\hat{M}_{k}$. 
\textcolor{black}{One can integrate over the fermionic degrees of freedom and find the mean field equations as in Ref. \cite{Kloss151}:} 
\begin{align}
&\color{black}{\hat{M}_{\omega,\bf{k}}=T\sum_{\omega',\bf{k}'}J_{\bf{k-k'-Q}}\hat{G}_{\omega',\bf{k}'}.}
\label{MFeq}
\end{align}
\textcolor{black}{The matrix equation can be projected onto the different components. The exact solution of these equations goes beyond the goal of the paper. However, we directly see that the order parameters depend on the frequency as well as on the momentum.}
\end{subsection}
\begin{subsection}{\textcolor{black}{Frequency dependence at the hot-spot}}
\textcolor{black}{The frequency dependence is a key player of the theory. Then we solve the equation (\ref{MFeq}) for the superconducting component $\Delta_{\omega,\bf{k}}$ only, putting CDW order parameter to zero ($\Delta^{CDW}_{\omega,\bf{k}}=0$). Considering a temperature of $T=0.001$ and a paramagnon mass to $m=10^{-6}$ with $\gamma=10^{-5}$ and $\lambda=12$, the frequency dependence of the SC order parameter at the hotspot is shown in Fig. \ref{freqdep}.}

\begin{figure}
\includegraphics{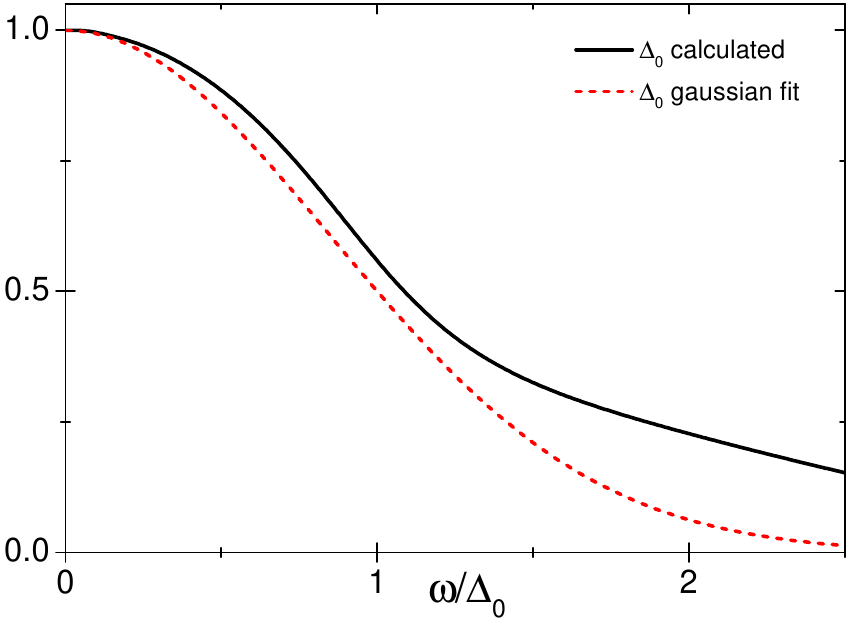}
\caption{\label{freqdep} \textcolor{black}{The frequency dependence of the superconducting order parameter $\Delta_{\bf{k}}$ at the hot-spot calculated from the mean field equation (\ref{MFeq}). The frequency form factor which is a Gaussian with a half-width of $\Delta_{0}$ which is the value of the SC order parameter at zero frequency.}}
\end{figure}

\textcolor{black}{As shown in Fig. \ref{freqdep}, the Gaussian frequency form factor can reproduce the frequency dependence of the SC order parameter. In the main text, we assume that the frequency dependence of the CDW order parameter can be fit by the same form factor. The calculation of the frequency dependence of the order parameters at each doping and each temperature is left for later work. }

\end{subsection}
\end{section}
\begin{section}{Raman experiments}
\label{RamanExperiment}
Raman experiments were carried out using a triple grating spectrometer (JY-T64000) equipped with a
liquid-nitrogen-cooled CCD detector. The 532 nm laser excitation line was used from respectively a diode pump solid state laser. Measurements in the SC state have been performed at $10\,K$ using an ARS closed-cycle He cryostat. Raman study was performed on Bi 2212 single crystals with two distinct levels of doping $p=0.16$ ($T_{c}=90\,K$) and $p=0.23$ ($T_{c}=52\,K$). The level of doping was controlled only by oxygen insertion (see Ref. \cite{sacuto15}). The $B_{1g}$ channel was obtained from crossed polarizations at $45^{0}$ from the Cu-O bond direction. The $A_{1g}$ channel was obtained from parallel polarizations along the Cu-O bond (given $A_{1g}$+ $B_{1g}$) and normalized subtraction of the $B_{1g}$ channel. All the spectra have been corrected for the Bose factor and the instrumental spectral response. They are
thus proportional to the imaginary part of the Raman response function Im $\chi_{R}^{\lambda} (\omega)$. ($\lambda=A_{1g}$ or $B_{1g}$)
\end{section}
\end{appendix}
\newpage
\bibliographystyle{apsrev4-1}
\bibliography{Cuprates}
\end{document}